\documentclass[11pt,titlepage]{article}

\usepackage{setspace} 
\usepackage{subfigure} 

\usepackage{amsmath,enumerate}

\usepackage{graphicx}

\usepackage[round,numbers,sort&compress]{natbib} 
\usepackage{color}
\usepackage{overpic}
\title{Structural Rigidity of Paranemic (PX) and Juxtapose (JX) DNA Nanostructures}

\author{
\bf{Mogurampelly Santosh}\\
Center for Condensed Matter Theory, Department of Physics,\\
Indian Institute of Science, Bangalore, India 560012.\\
E-mail: \it{santosh@physics.iisc.ernet.in}\\
\and \\
\bf{Prabal K. Maiti}\footnote{Corresponding author. Address: 
           Center for Condensed Matter Theory,
           Department of Physics, Indian Institute of Science,
	   Bangalore, India, 560012.
	   E-mail: maiti@physics.iisc.ernet.in,
	   Tel.:~(0091)80-22932865} \\
Center for Condensed Matter Theory, Department of Physics,\\
Indian Institute of Science, Bangalore, India 560012.\\
E-mail: \it{maiti@physics.iisc.ernet.in}
}
\date{}

\begin{document}

\maketitle

\abstract{
Crossover motifs are integral components for designing DNA based nanostructures
 and nanomechanical devices due to their enhanced rigidity compared to the normal B-DNA.
 Although the structural rigidity of the double helix B-DNA has been investigated
 extensively using both experimental and theoretical tools, to date there is no 
quantitative information about structural rigidity and the mechanical strength of
 parallel crossover DNA motifs. We have used fully atomistic molecular dynamics 
simulations in explicit solvent to get the force-extension curve of parallel DNA 
nanostructures to characterize their mechanical rigidity. In the presence 
of mono-valent Na$^+$ ions, we find that the stretch modulus ($\gamma_1$) of 
the paranemic crossover (PX) and its topo-isomer JX DNA structure is significantly 
higher ($\sim$ 30 \%) compared to normal B-DNA of the same sequence and length.
 However, this is in contrast to the original expectation that these motifs are 
almost twice rigid compared to the double-stranded B-DNA. When the DNA motif is 
surrounded by a solvent with Mg$^{2+}$ counterions, we find an enhanced rigidity 
compared to Na$^+$ environment due to the electrostatic screening effects arising 
from the divalent nature of Mg$^{2+}$ ions. This is the first direct determination 
of the mechanical strength of these crossover motifs which can be useful for the 
design of suitable DNA for DNA based nanostructures and nanomechanical devices 
with improved structural rigidity.\\

}

\clearpage

\section{Introduction}
The structural properties that enable DNA to serve so effectively as 
genetic material can also be used for DNA based computation and for 
DNA nanotechnology \citep{seeman_nature1}. The success of these 
applications to large extent, depends on the complementarity 
that leads to the pairing of the strands of the DNA double helix. 
This pairing can be exploited to assemble more complex motifs, 
based on branched structures with sticky ends that are promising 
for building macromolecular structures \citep{seeman_nature1,seeman_nl}. 
The combination of sticky ended ligation \citep{cohen} and stable 
branched DNA species \citep{seeman_jtb} has permitted investigators 
to construct DNA molecules whose edges have connectivity of a 
cube \citep{seeman_nature5} and of a truncated octahedron
 \citep{seeman_jacs1}. Similarly, branched DNA molecules have 
provided the basis for the syntheses of variety of structures such as 
knots \citep{seeman_jacs2,seeman_biopol}, Borromean rings 
\citep{seeman_nature4} and DNA tweezers \citep{simmel}. 
One of the early goals of DNA nanotechnology was to construct
 precisely configured materials on a much larger scale
 \citep{seeman_nature1, markus_2}. However the 
key feature that was lacking in such constructions were rigid 
DNA molecules, as flexible components failed to maintain the 
same spatial relationships between each member of a set 
\citep{seeman_nanotech}. DNA anti-parallel double crossover 
molecules (DX) \citep{seeman_bc}, analogous to intermediates
 in genetic recombination, were found to be about twice as rigid 
as a normal DNA molecules \citep{seeman_biophysj} and were subsequently
 used for creating two dimensional DNA lattices \citep{seeman_nature3}.\\

The concept of double crossover molecules stems from the concept of 
formation of Holliday junction \citep{holliday}, which is the most prominent 
intermediate in genetic recombination. These crossover molecules 
usually contain two Holliday junctions connected by two double helical 
arms. Depending on whether the crossovers are between strands of the 
same (called parallel) or opposite (called anti-parallel) polarity, 
as well as the number of double helical half-turns (even or odd) 
between the two crossovers, different double crossover molecules are 
generated. It is known that anti-parallel double helical molecules 
are better behaved than the parallel ones and as mentioned above 
these anti-parallel molecules have found numerous applications in 
nano construction \citep{seeman_jacs4}. The concept of crossover 
molecules also led to the discovery of paranemic crossover molecules (PX) 
and its topoisomers, the various JX molecules. The PX motif arises from 
the fusion of two parallel double helices by reciprocal exchange at 
every possible point where the two helices come in contact, whereas 
its topoisomer JX$_{i}$ contains {$i$} adjacent sites where backbones 
juxtapose without crossing over \citep{seeman_jacs3}. The rigidity of 
these crossover motifs has led to wide applicability in DNA 
based nanotechnology \citep{seeman_nature2}.\\

Nevertheless, there is lack of quantitative knowledge about the 
rigidity of the paranemic PX DNA and its topoisomers JX DNA, 
even though they find applications in nano constructions, owing to 
their inherent rigidity. Using atomistic simulation studies on PX 
and JX structures we have shown that among all synthesized PX and 
JX DNA structures, PX-6:5 motif is the most stable and in comparison 
with JX motifs the rigidity order runs as PX-6:5 $> \text{JX}_1 > 
\text{JX}_2>\text{JX}_3>\text{JX}_4$ \citep{maiti1,maiti2}. Our 
predictions on thermodynamic calculations are in consistence 
with the recent experimental studies using calorimetric and denaturing 
gradient gel electrophoretic methods \citep{seeman_pxjx_thermodynamics}.
 In this report we make an attempt to get a quantitative measure of 
the rigidity of these structures from their force-extension curves. 
The concept that has been adapted is to pull these structures with 
an external force in constant-force ensemble \citep{santosh} and from 
the internal energy as a function of extension of the DNA motif, we 
obtain the stretch modulus using Hooke's law of elasticity. 
The stretch modulus in turn gives an estimate of the rigidity of the motifs.\\

Experimental and theoretical studies have established that when 
subjected to an external force, double-stranded DNA (dsDNA) 
exhibits different force-extension regimes, stemming from 
its unique double-helix structures \citep{bustamante1,ewald_artifact,
lavery1,rief}. For example, in the low force regime, the elasticity 
of dsDNA is entropy dominated and the experimental force-extension
 data obtained with the magnetic bead method can be described very 
well by the standard entropic worm-like chain model \citep{bustamante1,
vologodskii,marko}; in the high force regime (starting from 10 pN 
to 70 pN), which is accessible with optical tweezers or atomic force 
microscopy, when the external force is comparable with the basepair 
stacking interaction in dsDNA the polymers can be suddenly stretched 
to about 1.7 times its B-form length in a very narrow force range 
\citep{bustamante3, lavery2}. An explanation of this regime is 
attributed to the short-range nature of basepair stacking interactions. 
Typically in this regime, the increase in length of the DNA molecule 
is due to large distortion in the double helical structure resulting in a 
ladder-like structure which is termed as S-DNA \citep{lavery2,jie_nar2,jie_nar1,jie_pre2008} 
{where most of the H-bonds
 involving basepairing are intact. It has been shown 
that depending on the experimental conditions such as 
the force attachment and salt concentration, dsDNA 
can undergo either strand-unpeeling transition to ssDNA 
\citep{rouzina_bj2,mameren_pnas}, or transition to S-DNA
 where basepairs are intact (or B-to-S transition)
\citep{lavery2,jie_pre2004}.}
In the elastic regime of force-extension curve, 
DNA behaves similar to that of a spring and follows the simple Hooke's 
Law: $F = YA \Delta L/L$ where $Y$ is the Young Modulus and $A$ 
is the cross-sectional area of the DNA. The product of Young modulus 
and area, $Y\times A$ gives stretch modulus, $\gamma_1$. At large forces,
 the stacking potential can no longer stabilize the B-form conformation 
of dsDNA and the (optimally) stacked helical pattern is severely 
distorted \citep{btos}, and therefore a structural transition from 
canonical B-form to a new overstretched conformation. While all these 
features are well established for dsDNA and ssDNA, no theoretical or 
experimental studies exist for the force-extension behavior of cross-over 
DNA molecules such as DX, PX and various JX structures. This paper is 
the first attempt to address such lacunas. The paper is organized as 
follows: in the next section we describe building of PX/JX nanostructures 
and the simulation methodologies. In section \ref{results} we describe 
the results from our fully atomistic DNA pulling simulation and in 
section \ref{conclusions} we give a summary of the force-extension results and conclude.\\

\section{Computational Details and Methodology}
\subsection{Building PX/JX structures}
\begin{figure}
	\begin{center}
        \subfigure[]
        {
	\includegraphics[height=20mm]{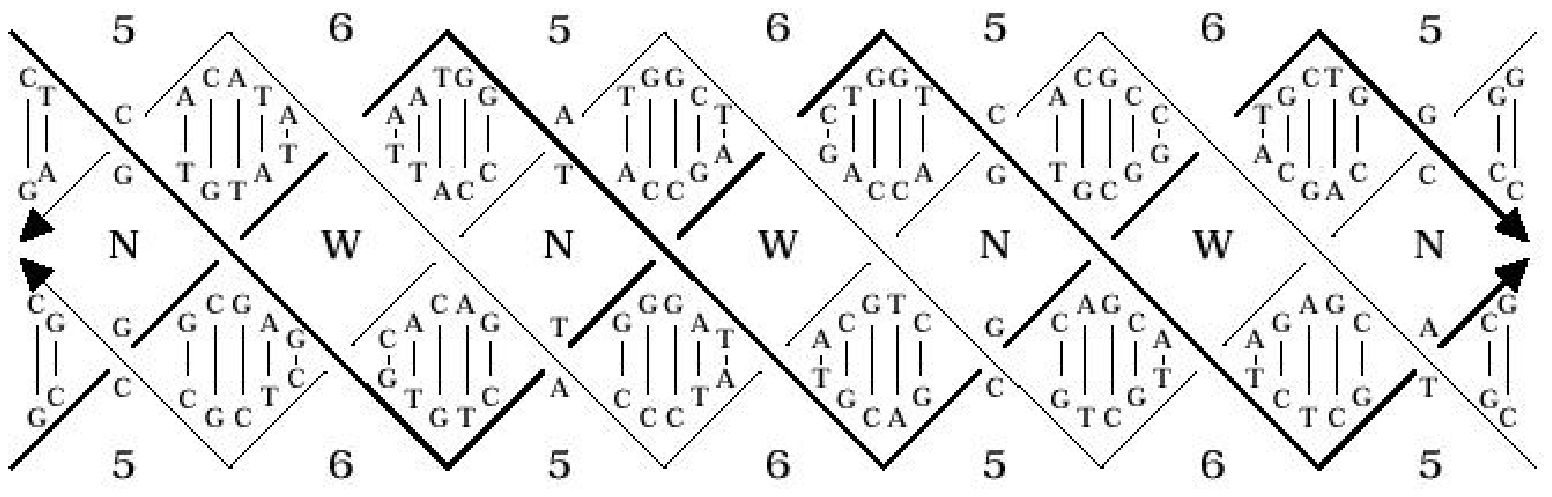}
	\label{bp-seq}
	} \\
        \subfigure[]
        {
	\includegraphics[height=50mm]{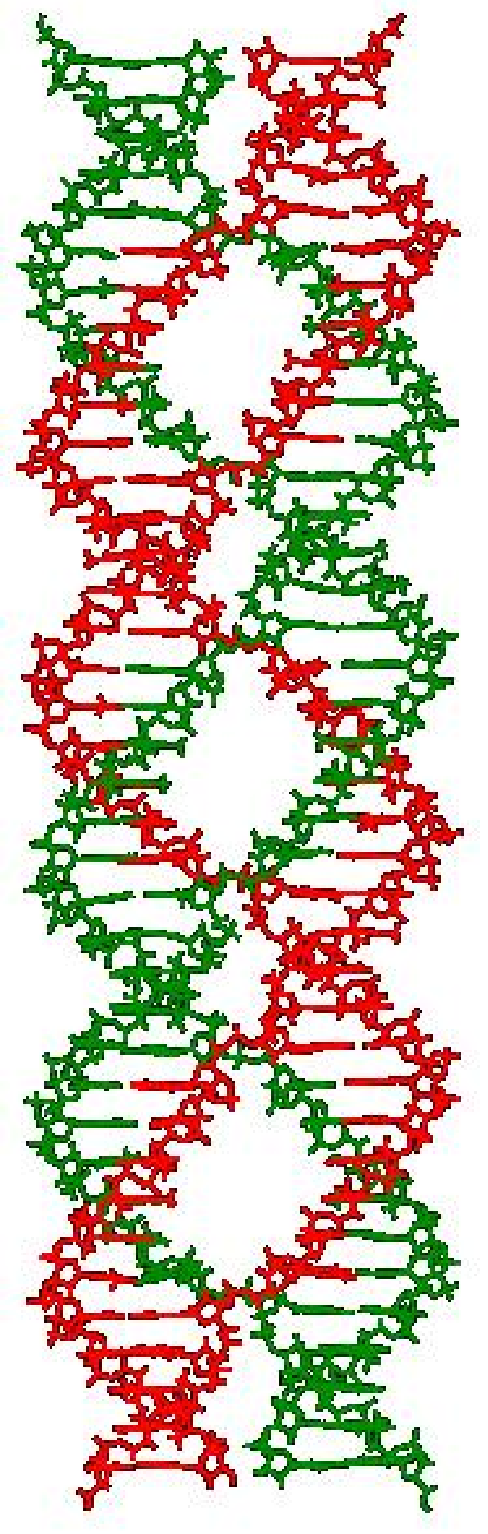}
	\label{init-str-px65}
	}
        \subfigure[]
        {
        \includegraphics[height=50mm]{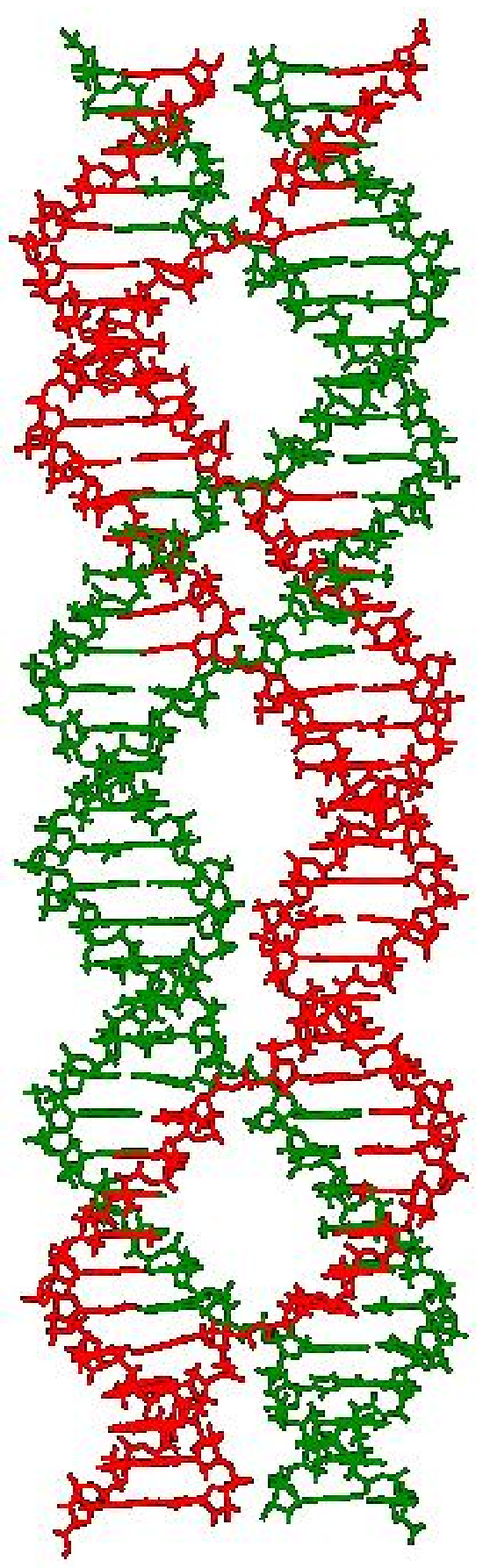}
        \label{init-str-jx1}
        }
        \subfigure[]
        {
        \includegraphics[height=50mm]{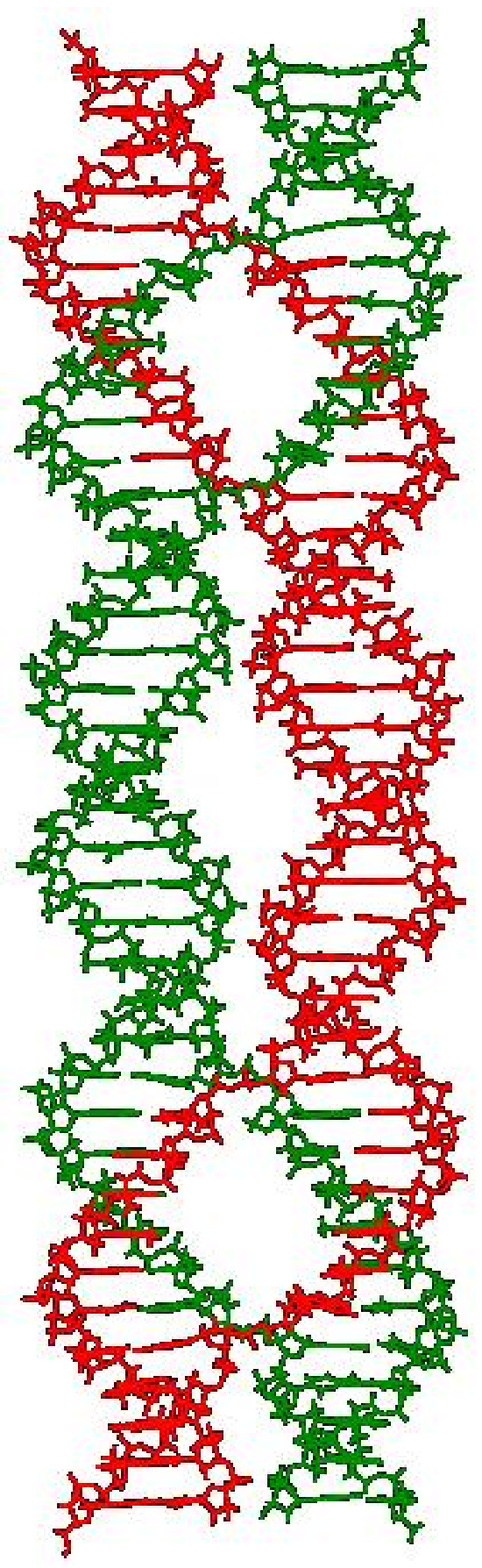}
        \label{init-str-jx2}
        }
	\caption{}
	\label{init-str}
        \end{center}
\end{figure}

In order to characterize the stiffness of various cross-over DNA motifs, 
theoretically we have tried to mimic nano-manipulation techniques like 
AFM, magnetic tweezers, or laser tweezers in our fully atomistic 
simulations. The basic protocol in our simulation technique was to put 
the DNA structure in counterion solution, keep one end of the structure 
anchored, and pull the other end along the PX molecule dyad axis 
(parallel to the helix axes) with an external force and determine the 
force-extension curves. The various DNA structures studied are PX-6:5, 
JX$_1$, JX$_2$, AT-6:5, GC-6:5 and normal B-DNA which are four-strand 
complexes of DNA paired with one another. Initial structures of PX-6:5, 
JX$_1$ and JX$_2$ were shown in Fig. \ref{init-str}. 
We have also studied the force-extension behavior of a double strand 
DNA having the same length (38 bps) and the sequence of one of the 
helices of the PX structure. This will allow us to compare the 
force-extension behavior of the normal double strand B-DNA with 
that of the cross-over DNA nanostructures. 
{The procedure for 
constructing these structures is as follows:
\begin{enumerate}[\hspace{0.1cm}a.]
\item{\emph{Building the DNA double helices:} Regular B-DNA has between
 10 and 10.5 base pairs per helical turn. Hence by varying the 
twist angle of a selected number of base pairs, we can create 
B-DNA structures with between 8 and 18 base pairs per helical turn. 
Table 1 of \citep{maiti1} shows the twist angles used for building the various PX 
structures. We assigned the same twist angle for all the base 
pairs in the helical half turn. The helical rise value of 3.4 \AA~
 was used to build the PX structures.}
\item{\emph{Building the crossover points:} When a double helix
 is built in Namot2 (version 2.2.) \citep{namot}, 
the molecules are oriented so that the 5$^\prime$
 and 3$^\prime$ ends of the double helices are parallel to the 
y-axis. To create realistic crossover structures, it is necessary 
to rotate the individual helices so that the desired crossover
 points are closest to each other (rotation angles shown in 
Table 2 of \citep{maiti1}). To find this point we wrote a program that starts
 with the first crossover point and rotates the first helix in 
1$^\circ$ increment to find the rotation leading to the 
shortest distance between these crossover points. Once found,
 the first helix is rotated by the prescribed value and held 
steady while the second helix is rotated and the shortest
 distance between the crossover points determined. The second
 helix is rotated 180$^\circ$ more than the 1st helix so that 
the helices are arranged as shown in Fig \ref{init-str}.  The crossovers
 were then created using the ``nick" and ``link" commands in Namot2.
 These structures are saved in the Protein Database (PDB) file format.}
Fig. \ref{init-str} shows the snapshot of the built PX, JX$_1$ and
 JX$_2$ structures after minimizations.
\end{enumerate}}

\subsection{Simulation Details for the PX/JX structures}
All molecular dynamics simulations were carried out using AMBER8 
suite of programs \citep{amber8} using the Amber 2003 (ff03) force 
fields \citep{ff03,kollman} and the TIP3P model \citep{tip3p} 
for water. For Mg$^{2+}$ we have used the \AA qvist \citep{aqvist}
 interaction parameters. This initial crossover DNA motif is 
then solvated with TIP3P model \citep{tip3p} water box using 
the LEaP module in AMBER8 \citep{amber8}. In addition, 
some water molecules were replaced by Na$^{+}$ (Mg$^{2+}$) 
counterions to neutralize the negative charge on the phosphate 
backbone groups of the PX/JX DNA structure. 
{We have used \AA qvist parameter \citep{aqvist}
 set to describe the ion-water as well as ion-DNA interactions.
\AA qvist \citep{aqvist} parameters reproduce the location of the first peak 
of the radial distribution function for the corresponding 
ion-water and the solvation free energies of several ionic species.
After neutralizing the system with counterions,} 
the concentration of Na$^+$ (Mg$^{2+}$) 
in the PX-6:5, JX$_1$, JX$_2$, AT-6:5, GC:6-5 and normal B-DNA 
crossover systems is 246 mM (160 mM) where as for a double 
helix B-DNA, the concentration of Na$^+$ (Mg$^{2+}$) is 254 mM (140 mM). 
The ion specificity and electrostatic interactions play crucial in the 
DNA functioning and protein-DNA binding mechanism \citep{nathan_pnas2001, nathan_2005}.
{The LEaP module works by constructing Coulombic potential 
on a grid of 1 \AA~ resolution and then placing ions one 
at a time at the highest electrostatic potential.
Once the placements of all the ions are done using 
the above method long MD simulations ensure that they
 sample all the available space around DNA and preferentially
 visit electronegative sites. In fact the initial placement 
of ions should not influence the final results provided long
 simulation was performed before the pulling runs were done.
 The radial distribution function provided in the Fig. S1 
and S2 of supplementary material also reveal this fact that the 
counterions remain associated in the close proximity
 ($\sim$ 10 \AA) of the DNA. We have also made sure 
that no counterion is ``stuck" to the DNA molecule 
anywhere and observed diffusive behavior. Henceforth 
we believe that the initial ion distribution is 
important to study DNA properties but ultimately 
equilibration is the key for avoiding such initial 
ion position dependency.}
The box dimensions were chosen in order to ensure a 10 \AA~ 
solvation shell around the DNA structure in its fully extended 
form when the DNA is in overstretched regime. This procedure 
resulted in solvated structures, containing 99734 atoms for 
PX-6:5, JX$_1$ and JX$_2$; 97326 atoms for 38-mer B-DNA; 99781 
atoms for PX-6:5 structure having only AT sequence and 99705 
atoms for PX-6:5 structure having only GC sequence in a box of 
dimensions $54\times77\times299 \text{~\AA}^3$ when Na$^+$ ions
 are present as counterions. Translational center of mass 
motions were removed every 1000 MD steps. The trajectory was 
saved at a frequency of 2 ps. We have used periodic boundary 
conditions in all three directions for the water box during the 
simulation. Bond lengths involving bonds to hydrogen atoms were 
constrained using SHAKE algorithm \citep{shake}. During the 
minimization, the PX/JX structure was fixed in their starting 
conformations using harmonic potential with a force constant of 
500 kcal/mol-\AA$^2$. This allowed the water molecules to reorganize 
to eliminate bad contacts within the system. The minimized 
structures were then subjected to 40 ps of MD, using 1 fs time 
step for integration. During the MD, the system was gradually heated 
from 0 to 300 K using weak 20 kcal/mol-\AA$^2~$ harmonic 
constraints on the solute to its starting structure. 
 NPT-MD was used to get the correct solvent 
density corresponding to experimental value of the density. 
Lastly, we have carried out pulling the motif in NVT MD with 2 
fs integration time step using a heat bath coupling time constant of 1 ps.\\

The external pulling force was applied at one end on O3$^\prime$ 
and O5$^\prime$ atoms on each strand and the other O5$^\prime$ and 
O3$^\prime$ atoms of the strands on the other end were fixed with 
large force constant of 5000 kcal/mol-\AA$^2$. The force was 
applied along the direction of the end-to-end vector joining 
O3$^\prime$ and O5$^\prime$ atoms. The external force started 
at 0 pN and increased linearly with time 
up to 1000 pN with a rate of force of 10$^{-4}$ pN/fs. 
 For comparison, we have also done the simulation with faster
pulling rate at 10$^{-3}$ pN/fs. It should be mentioned that
the typical pulling rate in an AFM experiment is of the order
of 10$^4$ pN/s whereas the slowest pulling rate achieved in our
 simulation is 10$^{11}$ pN/s (or 10$^{-4}$ pN/fs) due to
computing limitations. Theoretical model suggests that with the
increased rate of force, the dynamic strength of the molecular
adhesion bonds increases \citep{evans}. Therefore, since our simulation
employs higher rates of force compared to AFM pulling rates, the
calculated stretch modulus will be higher than that of the stretch
modulus calculated in an AFM pulling experiment. 
Our pulling protocol was validated for B-DNA \citep{santosh} 
and is expected to be applicable for PX/JX motifs also. 
In Fig. S3(a) we have shown the 
force-extension curve that is consistent with experimental 
curve. Note that the plateau region was observed at 95 pN 
instead of 65 pN which has its origin in fast rate of pulling 
which is inherent to computer simulations. This MD simulation 
has been done with a pulling rate of 10$^{-5}$ pN/fs, the 
slowest pulling rate we could achieve as of now. 
{We have also done the pulling
simulation at various rates. In Fig. S3(b) we plot the 
plateau force as a function of the pulling rate. 
From this plot we can see that by extrapolating to 
slower pulling rates, one can get a plateau region 
like that of AFM pulling experiments. Hence} 
we expect that we can exactly get the
 experimental force-extension curve 
when pulled much slower than 10$^{-5}$ pN/fs such as in AFM or 
optical tweezers. {Apart from the rate dependency,} 
the stretch modulus calculation is in 
accordance with the experimental result.\\

\section{Results and Discussion}
\label{results}
\subsection{Force-extension behavior}
\begin{figure}
        \begin{center}
        \subfigure[]
        {
    \begin{overpic}[scale=0.634]{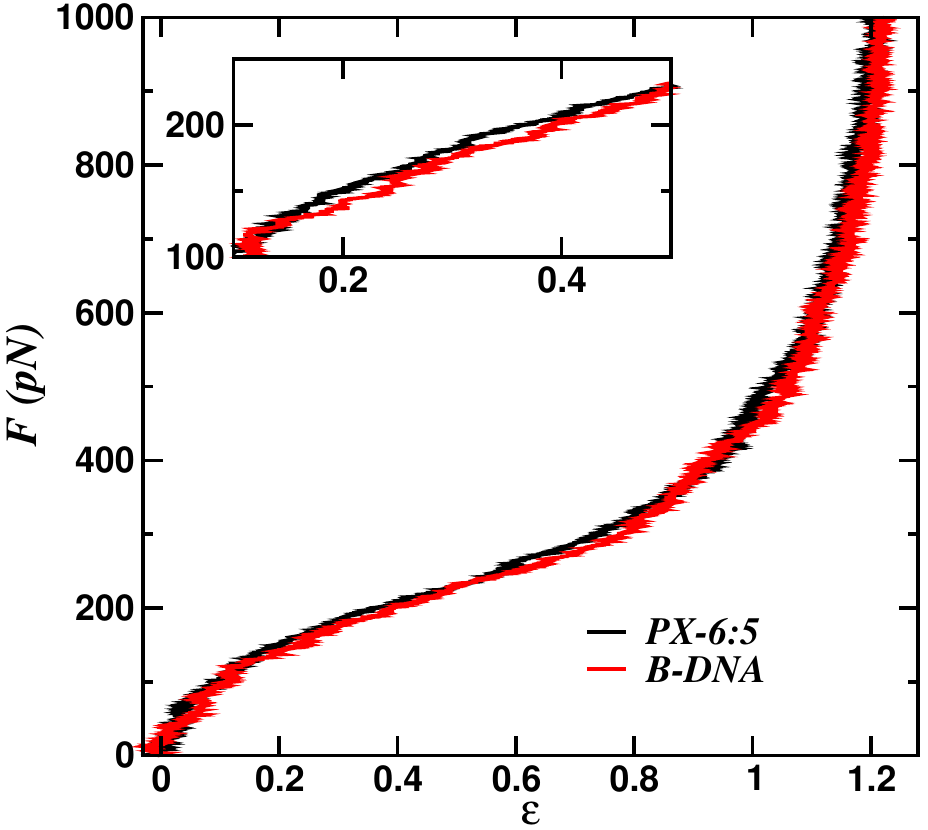}
    \put(38,23){$\uparrow$}
    \put(21,27){\includegraphics[scale=0.15]{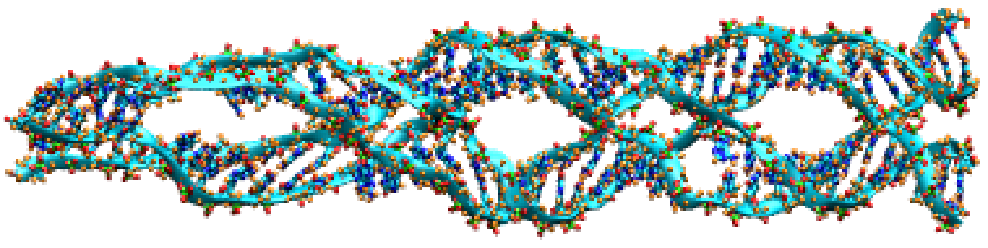}}
    \put(68,38){$\leftarrow$}
    \put(31,38){\includegraphics[scale=0.15]{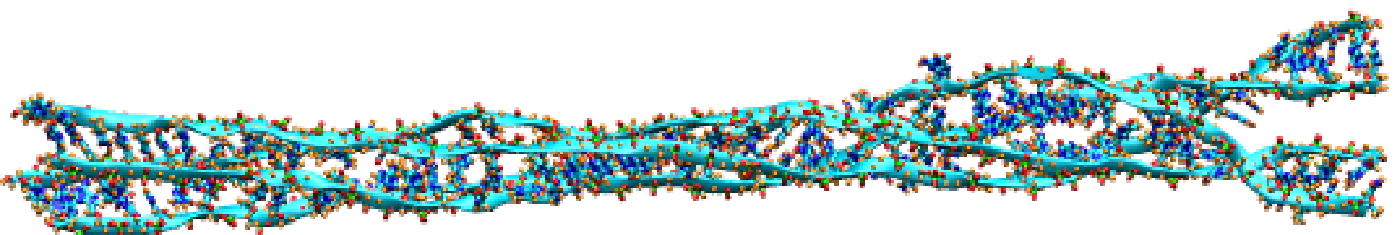}}
    \put(81,60){$\swarrow$}
    \put(40,48){\includegraphics[scale=0.15]{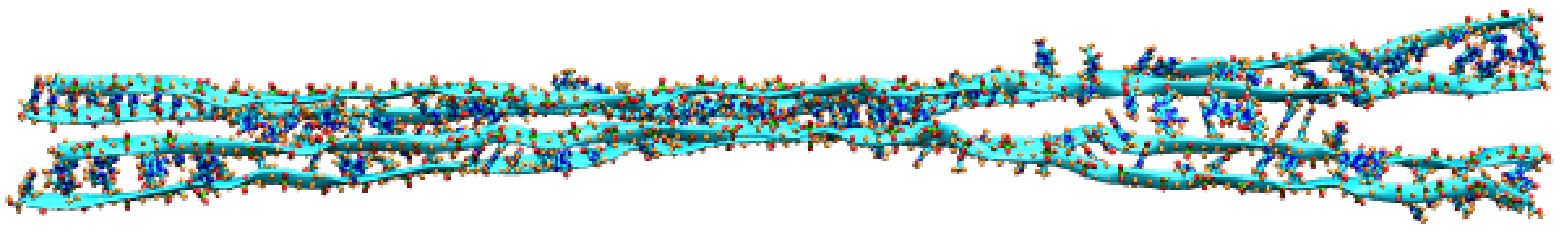}}
    \end{overpic}
	\label{force-ext-px}
	}
        \subfigure[]
        {
        \includegraphics[height=53mm]{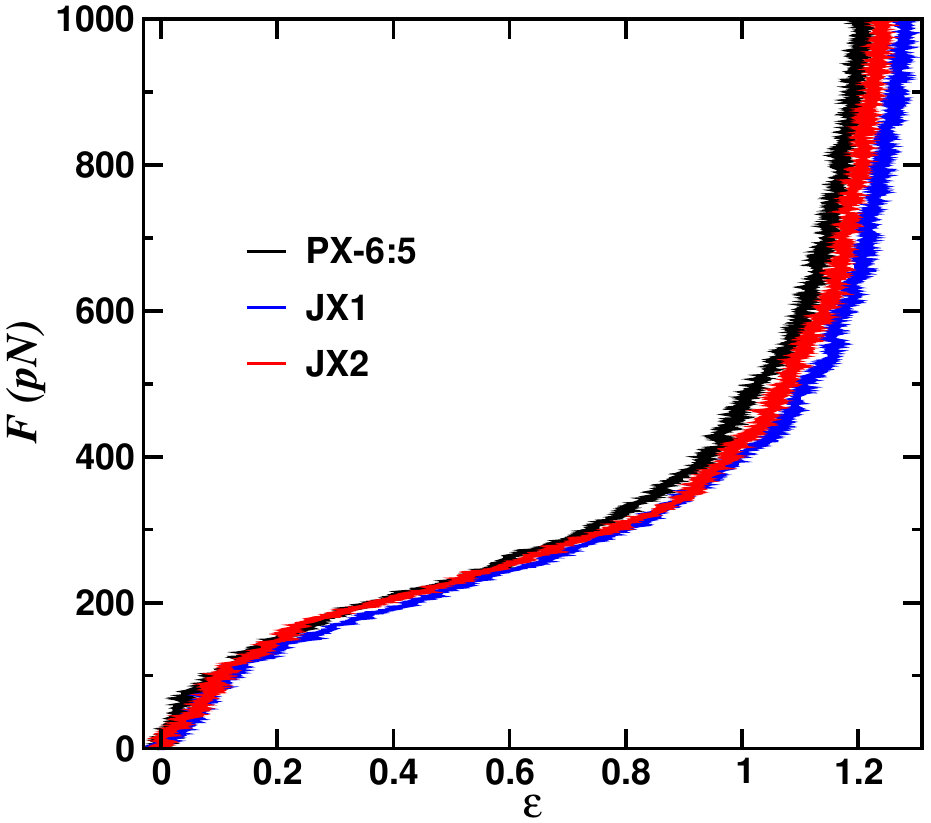}
        \label{force-ext-all}
        }
	\caption{}
	\label{force-ext}
        \end{center}
\end{figure}

\begin{figure}
        \begin{center}
        \includegraphics[height=80mm]{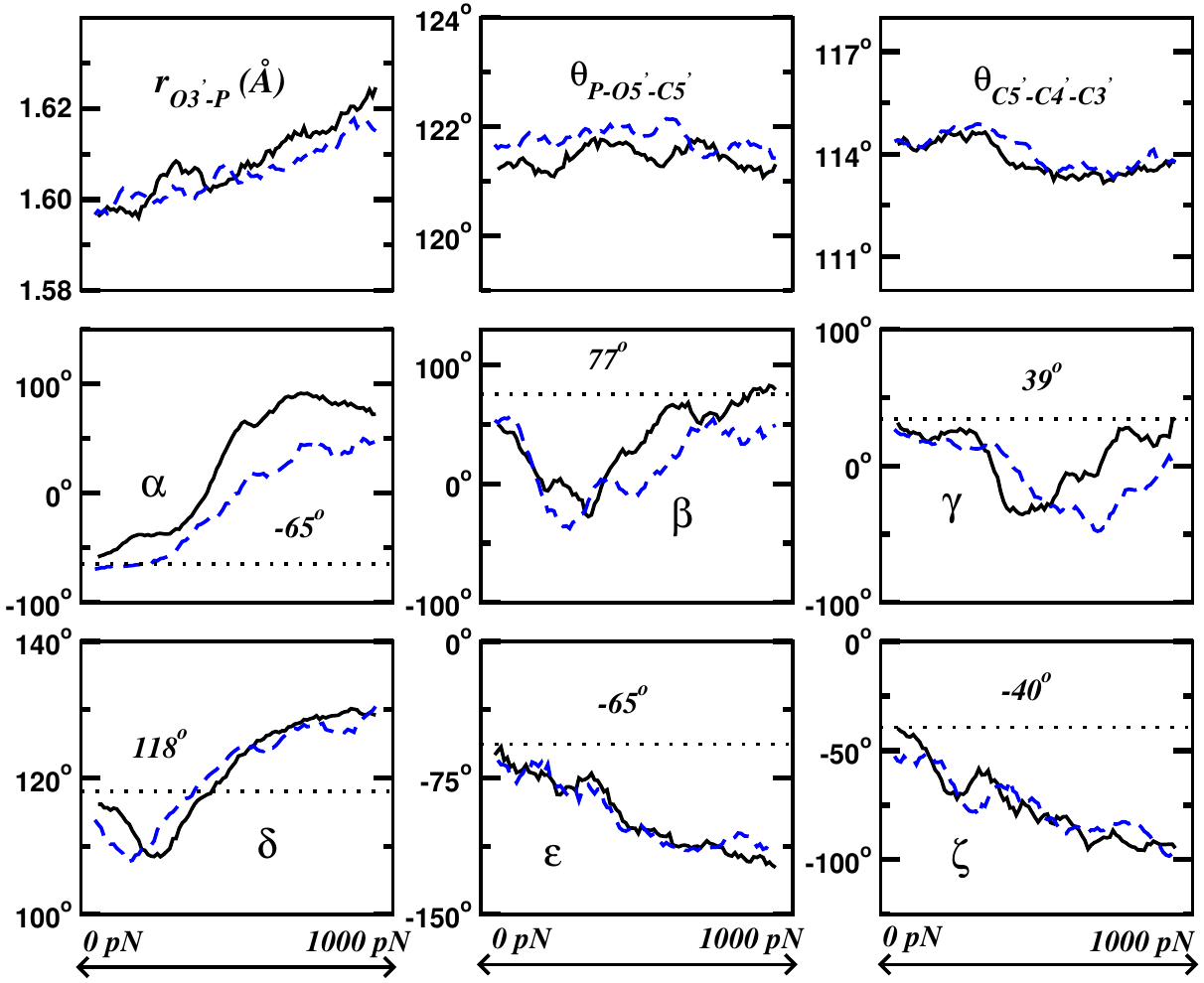}
        \caption{}
        \label{dist_angle_torsion}
        \end{center}
\end{figure}

Fig. \ref{force-ext} gives the force-extension curve for PX-6:5 
DNA motif in the presence of Na$^+$ counterions. The force-extension 
curve consists of an entropic regime where, the extension of DNA beyond 
its contour length is negligible and this regime continues until 150 pN. 
{Beyond this forces and up to 200 pN, DNA extends to 
about 30 \% with most of the H-bonds are still intact. It has been 
shown that depending on the experimental conditions such as 
the force attachment and salt concentration, dsDNA 
can undergo either strand-unpeeling transition to ssDNA
\citep{rouzina_bj2,mameren_pnas}, or transition to S-DNA
 where basepairs are intact (or B-to-S transition) 
\citep{lavery2,jie_pre2004}. 
This is followed by an overstretching plateau region
 resulting in an extension of 1.7 times
the initial contour length where the DNA helical structure starts to
deform.}
Beyond this plateau regime is the overstretched structure 
of DNA which is followed by strand separation. It is worth 
mentioning that at low rates of force, the above 
mentioned regimes in the force-extension curves shifts to much lower 
force values. For comparison we have also shown the force-extension 
behavior of the normal double strand B-DNA which has same length and
 sequence as one of the helices of the PX-6:5 structure. From the 
force-extension curve we can calculate the stretch modulus of DNA in 
the elastic regime using Hooke's Law. However, it is known from previous
 theoretical and experimental studies on the DNA force-extension behavior 
that at high force regimes a conventional WLC model and its variant does 
not reproduce the force-extension behavior very well. Previous theoretical 
studies have demonstrated that at high force regime, non-linear 
elasticity plays an important role governing the force-extension
 behavior of the DNA \citep{netz} and the WLC model is inadequate 
to describe the force-extension behavior 
\citep{nelson_bj2007,nelson_naturenanotech2006}. 
Therefore, we have also used the following polynomial function to fit 
the simulated force-extension curve for whole force regime
\begin{equation}
F = \sum_{n=1}^{\infty}\gamma_n\left(\frac{L}{L_0}-1\right)^n = \sum_{n=1}^{\infty}\gamma_n \varepsilon^n
\end{equation}
Integrating the above equation, we get energy
\begin{equation}
E = E_0 + L_0 \sum_{n=2}^{\infty}\frac{\gamma_{n-1}}{n}\left(\frac{L}{L_0}-1\right)^n = E_0 + L_0 \sum_{n=2}^{\infty}\frac{\gamma_{n-1}}{n} \varepsilon^n
\label{eqn_energy}
\end{equation}
Where $E_0$ is the energy of the DNA when the extension is zero, 
$L$ is the length of the DNA and $L_0$ is the initial contour length of the DNA. 
The coefficients $\gamma_{n}$ for various values of $n$ gives various elastic modulii. 
For example, $\gamma_{1}$ gives the linear stretch modulus. 
Stretch modulus $\gamma_{1}$ can be calculated from both the above 
equations by fitting the force and energy as a function of the applied strain. 
From the above two equations, it can be seen that the linear term 
coefficient and the quadratic term coefficient will give the stretch 
modulus, respectively. We have done a sequential fit to the energy vs strain 
curve similar to ref \citep{netz} where the fitting is done for the
 leading quadratic form of energy vs strain plot for restricted data 
that are taken around the energy minima and obtain the stretch modulus 
as a coefficient of the quadratic term. 
{By giving more weightage to the quadratic term,
we have done the higher order fit due to the difficulty 
in identifying the exact Hooke's law region.
The term `sequential' essentially means
 that the leading terms near the energy
minima are fit to quadratic term and the rest 
are fit to higher order terms. This procedure is repeated self
consistently so that the value of $\gamma_1$ 
is independent of the $\gamma_n$ for large $n$.
However $\gamma_1$ depends very weakly on $\gamma_2$
 and $\gamma_3$ (that means $\gamma_1$ has lesser 
dependence on choice of $\gamma_n$ for $n>1$). 
The energy vs strain curve has a minimum around 
which a quadratic fitting was done and the obtained $\gamma_1$ is
the stretch modulus as shown in equation \ref{eqn_energy}.}
The stretch modulus obtained
from a polynomial fit to the force-extension
 curve and also from a sequential fit to the energy-strain 
curve are listed in table 
\ref{table1}.
 The values of stretch modulus from the above two fitting methods 
are very similar. From the sequential 
fit to the energy-strain curve, we find that PX-6:5 has a stretch modulus 
of 1636 pN at 246 mM of Na$^+$ concentration when pulled with 10$^{-4}$ pN/fs 
rate, which is 30 \% more rigid compared to the stretch modulus 1269 pN of
 B-DNA at same concentration and of the same sequence. From the 
polynomial fit to the force-extension curve, the same trend was 
observed (Table \ref{table1}). 
{During the pulling the bond lengths and 
bond angles (Fig. \ref{dist_angle_torsion}) are changing very less whereas
the torsion angles (Fig. \ref{dist_angle_torsion}) are changing almost 100 \% 
with respect to the initial values at zero force
 (See next para for more discussion).
This implies that the backbone helix has large internal
 dynamics and offer less resistance to the
applied force. Note that the backbone helix gives the 
structural stability to the DNA molecule. On
the other hand, bond lengths and bond angles offer more 
resistance to the applied force. Hence the
crossover links between helical strands contribute
 more to the structural rigidity of these molecules.
Although 30 \% of rigidity increase seems small, 
yet this enables one to construct promising DNA
nanotechnology devices with enhanced rigidity.} 
For the JX$_1$, JX$_2$ topoisomers at the same 
counterion concentration we obtain 1515 pN, 1349 pN 
from sequential fitting and 1373 pN, 1521 pN from polynomial fitting 
respectively. Force-extension plot for PX-6:5, JX$_1$ and JX$_2$ is shown 
in Fig. \ref{force-ext-all} and can see a slight change in the slope of 
linear region. Among all structures, PX-6:5 has large stretch modulus 
both in Na$^+$ and Mg$^{2+}$ medium due to the most number of crossover points. 
It was also shown that the optimal design of repeated stacks and bundles of 
nanostructures provide great strength to the molecule \citep{markus_buehler}. 
Experimental reports of stretch modulus for lambda phase DNA is of the order 
of 1000 pN depending on the environmental conditions like ionic 
strength \citep{bustamante1,bustamante2}. Though we get the stretch modulus 
of the same order, the helix to ladder transition occurs at a much larger 
force regime than that observed experimentally \citep{konrad,lavery3} due to 
the higher rates of force employed in our simulation. Experimental 
results have shown that the ladder transformation occurs at a much 
lower ($\sim$ 60-70 pN) force than that observed in our simulation data. 
It has been observed that a higher force rate leads to higher stiffness of 
the short DNA (Fig. \ref{rate}). This shift of curve and increase of slopes 
in force-extension curves is expected since higher rates means that our 
simulation is far from being reversible, in which case it dissipates energy 
to the environment (since our simulations were constant T = 300 K) and higher 
the irreversibility the higher is the work of dissipation which is given by 
the area under the force-extension curve. A possible reason may also be the 
length of DNA that has been used. Since a 38-mer DNA is too short, its contour 
length being significantly smaller than the persistence length of a lambda DNA, 
its behavior is likely to deviate from that observed for lambda DNA \citep{bustamante1,bustamante2}.\\

\begin{figure}
        \begin{center}
	\includegraphics[]{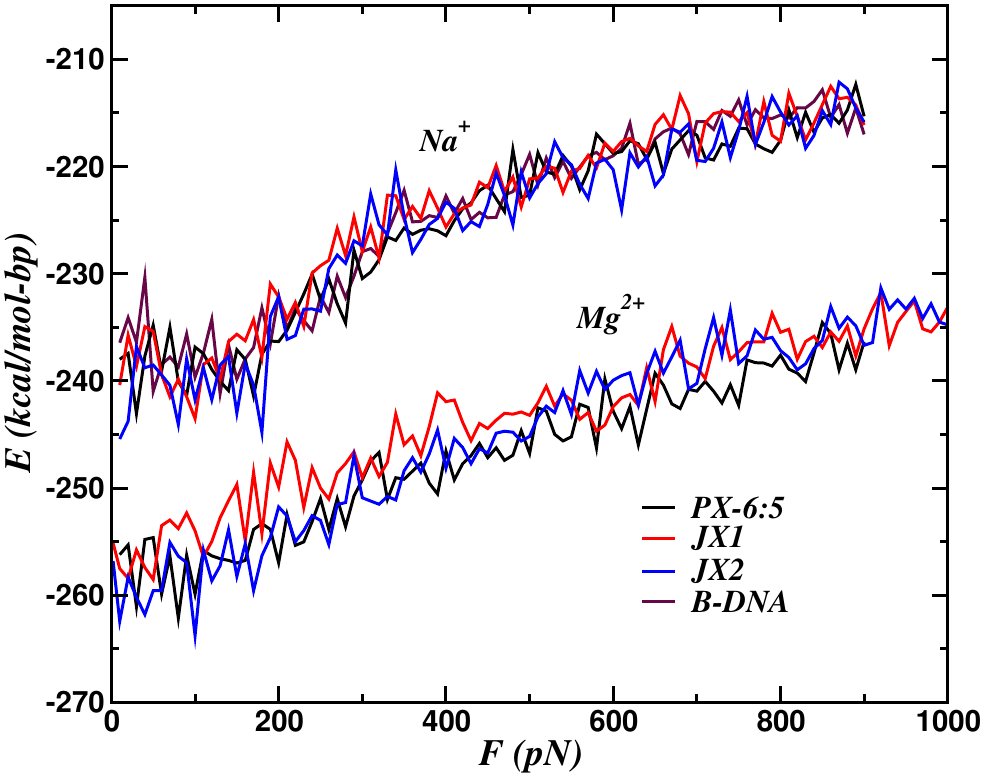}
	\caption{}
	\label{energy-force}
        \end{center}
\end{figure}

\begin{figure}
        \begin{center}
        \subfigure[]
        {
	\begin{overpic}[height=82mm]{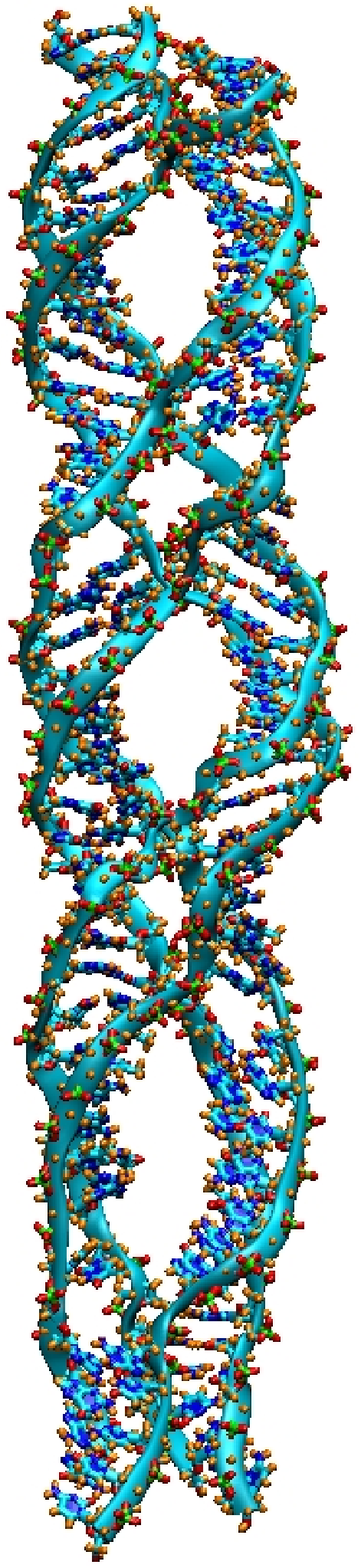}
	\put(5,107){\bf{200}}
	\put(5,102){\bf{pN}}
	\label{snapshots-px65-a}
	\end{overpic}
	}
        \subfigure[]
        {
	\begin{overpic}[height=82mm]{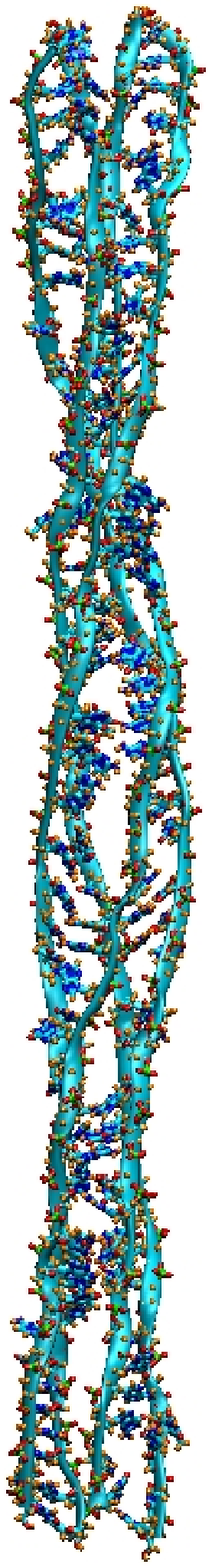}
	\put(2,107){\bf{400}}
	\put(2,102){\bf{pN}}
        \label{snapshots-px65-b}
	\end{overpic}
        }
        \subfigure[]
        {
	\begin{overpic}[height=82mm]{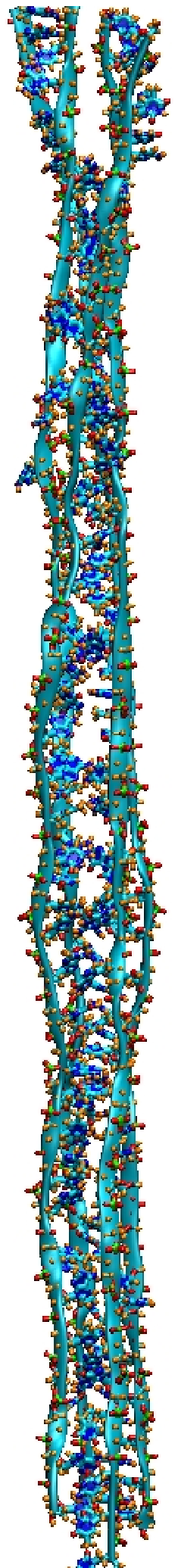}
	\put(0,107){\bf{600}}
	\put(0,102){\bf{pN}}
	\label{snapshots-px65-c}
	\end{overpic}
	}
        \subfigure[]
        {
	\begin{overpic}[height=82mm]{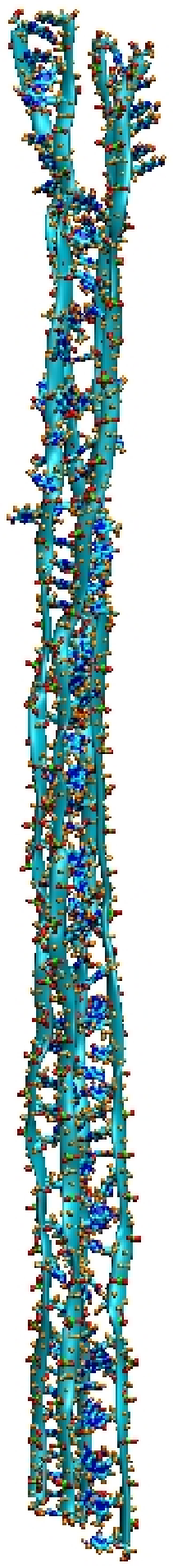}
	\put(1,107){\bf{800}}
	\put(1,102){\bf{pN}}
	\label{snapshots-px65-d}
	\end{overpic}
	}
        \subfigure[]
        {
	\begin{overpic}[height=82mm]{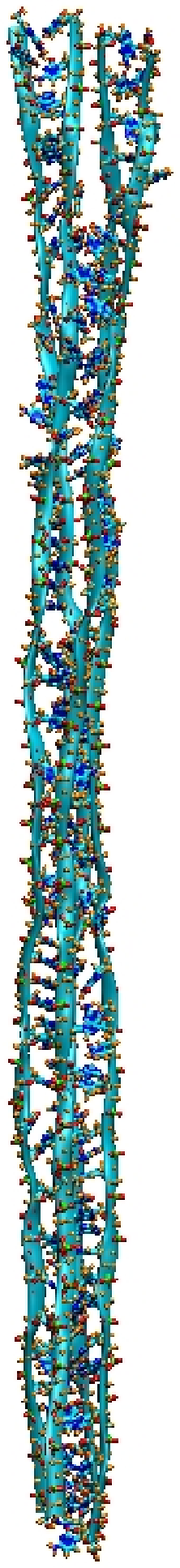}
	\put(0,107){\bf{1000}}
	\put(0,102){\bf{pN}}
	\label{snapshots-px65-e}
	\end{overpic}
	}
	\caption{}
	\label{snapshots-px65}
        \end{center}
\end{figure}

To gain further microscopic understanding of the structural
 changes with the applied force we have looked at the energetics
 of the PX/JX DNA as a function of the applied force. In Fig.
 \ref{energy-force} we plot the total internal energy as a function of 
force for PX-6:5, JX$_1$ and JX$_2$ DNA structures. The conformational 
entropy of the structure is dominating for small forces during which the 
change in length is almost negligible. For forces up to 150 pN, 
the energy is decreasing and seems to attain minimum value which corresponds 
to a most stable configuration. With further increase in force above 150 pN, 
the energy of the DNA molecule increases. This implies that the PX/JX DNA 
structures become thermodynamically unstable under elongation. 
Various backbone parameters were calculated when PX-6:5 structure is 
pulled at one end when the structure is neutralized with Na$^+$ and Mg$^{2+}$ counterions. 
The average bond length $\left(r_{\text{O3}^\prime-\text{P}}\right)$ and 
average angles $\left(\theta_{\text{P-O5}^{\prime}\text{-C5}^{\prime}}
\text{~and~}\theta_{\text{C5}^{\prime}\text{-C4}^{\prime}\text{-C3}^{\prime}}\right)$ 
were changing less than 2 \% with applied force with respect to the zero force 
equilibrium values (Fig. \ref{dist_angle_torsion}). Correspondingly, the increase in 
the energy contribution from bond stretching and angle bending is about to 5-10 \% 
compared to the zero force case. Interestingly, all the torsion angles 
($\alpha,\beta,\gamma,\delta,\epsilon,\zeta)$ \citep{saenger} were changing over 
100 \% (Fig. \ref{dist_angle_torsion}) with respect to the values at zero force, 
which is causing the structural deformation to great extent. The change in 
torsion angles is not considerable for forces up to 200 pN during which the 
change in DNA contour length is also very less. Dramatic change is observed in 
the torsion angles when the DNA motif elongates to almost twice of its initial 
contour length at critical force. At this stage, 
except the breaking of few H-bonds \citep{santosh}, 
we see no considerable change in average bond length or average angle. 
{We have used geometry measurement 
based criteria for the H-bond calculation
in simulation. Generally the H-bond 
is represented as D$\--$H$~^{\dots}$A where D is the donor and A
is the acceptor which is bonded to D through the H 
atom. In the case of DNA, D is a N atom and A
is either a N or O atoms depending on AT and GC 
base pairing. When the distance between D and A
atoms is less than 2.7 \AA~and the angle DHA is
 greater than 130$^\circ$, we say that the atom A is H-bonded 
to atom D otherwise the H-bond is broken.}
It is justifiable to argue that the backbone atoms in DNA motif were 
drastically re-oriented to give the elastic rod a sudden elongation, 
a clear signature indicating very large change in torsion angles.
 The torsion energy is also increasing greatly compared to all other 
contributions to the total internal energy of the motif. 
So a closer look at the various energy contributions reveals that, 
when the DNA motif is pulled along the helix axis, there is an increase of 
about 5-10 \% in the bond stretching and angle bending energy, 
15 \% increase in van der Waals energy but dramatic increase in the torsion energy. 
Apart from all the energy components like bond, angle, van der Waals, Coulomb energy, 
this increase in the torsion energy contributes to rapid increase in the total 
internal energy at a critical force.
Instantaneous snapshots of PX-6:5 
structure at different force in the presence of Na$^+$ counterions is 
shown in Fig. \ref{snapshots-px65} (see Fig. S4-S13 for instantaneous snapshots 
of PX-6:5, JX$_1$ and JX$_2$ in the presence of Na$^+$ and Mg$^{2+}$ 
at various forces, respectively). {
As the applied force is increased
beyond 200 pN, H-bonds in the basepair were broken and the DNA 
overstretches to 1.7 times its initial contour length.}\\

The original experiments on the PX/JX DNA molecules were performed in 
Mg$^{2+}$ buffer \citep{seeman_jacs3}. To understand the pulling response 
of these DNA nanostructures with divalent cations we have also done the
 pulling simulation of PX-6:5, JX$_1$ and JX$_2$ structure in the presence 
of divalent Mg$^{2+}$ ions. The counterion concentration in our simulation 
is close to 160 mM and the rate of pulling is 10$^{-4}$ and 10$^{-3}$ pN/fs.
{We have analyzed the force-extension spectrum 
of all DNA structures in Na$^+$ and Mg$^{2+}$ ions at 10$^{-4}$ pN/fs
(Fig. \ref{force_ext_spectrum}).}
Also, Fig. S14(a) shows the force-extension behavior of PX-6:5 molecule
 in presence of Mg$^{2+}$ ions. In the presence of Mg$^{2+}$, PX-6:5 have a 
stretch modulus of 1840 pN with a pulling rate of 10$^{-4}$ pN.
In contrast, the stretch modulus of B-DNA of same length and 
sequence as one of the helical domains of PX-6:5 is 1590 pN.
 Again we see the enhanced rigidity of the PX-6:5 
motif as was in the presence of Na$^+$ ions. However, this increase is not as 
dramatic as in the presence of Na$^+$ ions. Possibly the rigidity can be further 
enhanced by increasing the ionic strength. For all the structures we find the stretch 
modulus is more in the presence of Mg$^{2+}$ than in the presence of Na$^+$ counterions. 
{This is due to the strong phosphate-Mg$^{2+}$ 
coordination that resist the external applied force.} This result is in 
accordance with the experimental results on DNA stretch modulus 
\citep{bustamante2} where it is shown that the presence of divalent cations strongly 
reduce the persistence length of the DNA and hence increase the stretch modulus of 
the DNA. To understand the effect of crossover on the rigidity of the DNA motifs 
in the presence of Mg$^{2+}$ ions we have also done the pulling simulation 
for JX$_1$ and JX$_2$ motifs. Fig. S14(b) gives the 
force-extension behavior of the JX$_1$ and JX$_2$ motifs and 
compared with the PX-6:5 in presence of Mg$^{2+}$ ions. 
We get the stretch modulus from the sequential fitting of the energy vs strain plot: 
JX$_1$ and JX$_2$ have stretch modulus of 1465 pN and 1654 pN respectively. 
We see the same trend with both 10$^{-4}$ and 10$^{-3}$ pN/fs pulling rates as 
calculated from polynomial fitting. To get more of a molecular level picture we 
have plotted the internal energy of the PX/JX structure as a function of pulling
 force as shown in Fig. \ref{energy-force} and we see a similar behavior compared to 
the energy variation observed in the presence of Na$^+$ ions. Energy is decreasing in
 the small force regime thus making the DNA structures thermodynamically more stable. 
The increase in the energy with further load is rapid compared to the case of pulling 
in Na$^+$ medium. Increase in the total energy with the applied force implies that
 the PX/JX DNA structure is thermodynamically unstable, with some of H-bonds are broken.
We give all the stretch modulus results in both Na$^{+}$ 
and Mg$^{2+}$ counterion solution in Fig. S16.
 For comparison in Fig. S1 and S2 we show the radial distribution functions of 
the Na$^+$ and Mg$^{2+}$ ions with the O1P and O2P Oxygen of the phosphate backbone.

\subsection{Effect of DNA basepair sequence and rate of pulling}
To investigate the effect of sequence on the structural rigidity of the 
PX crossover DNA motifs we have calculated the force-extension profile for 
PX-6:5 made of only AT or GC basepairs in presence of Na$^+$ counterions. 
The stretch modulus obtained from the sequential fits comes out to be 1592 pN 
for AT sequence and 1780 pN for GC sequence. As expected, a PX structure made of
 only GC sequence is much stiffer than one made of only AT sequence due to extra
 H-bonding in GC basepair. The stretch modulus of crossover PX structures
 made of only AT or only GC is still larger than the stretch modulus of 1269 pN of
 B-DNA double helix of same length with a combination of AT and GC sequence. 
Stretch modulus $\gamma_1$ values for various PX-6:5 and JX structures are 
tabulated in table \ref{table1} for comparison.\\

\begin{figure}
        \begin{center}
        \subfigure[]
        {
	\includegraphics[height=60mm]{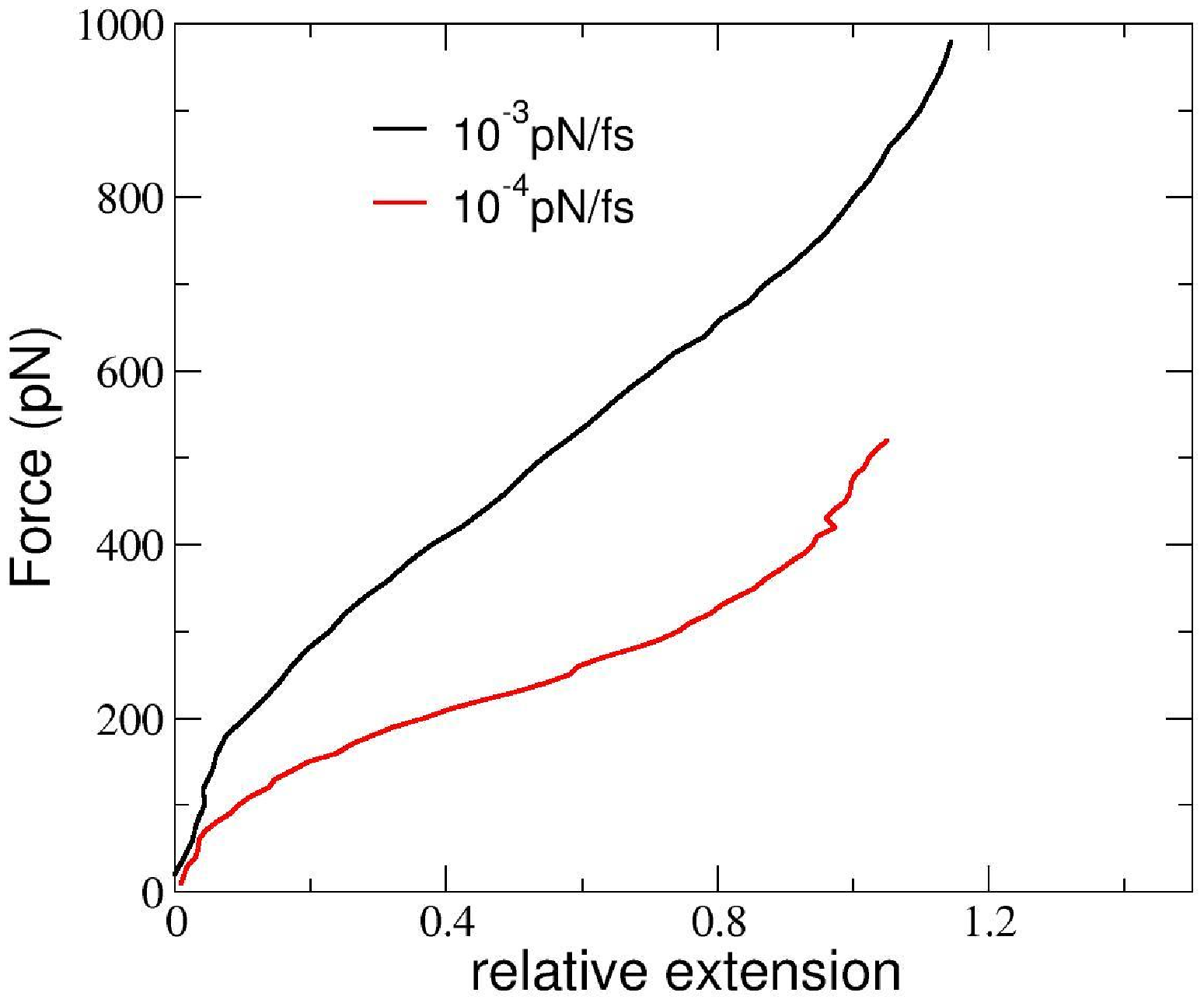}
	\label{rate-ext}
	}
        \subfigure[]
        {
        \includegraphics[height=60mm]{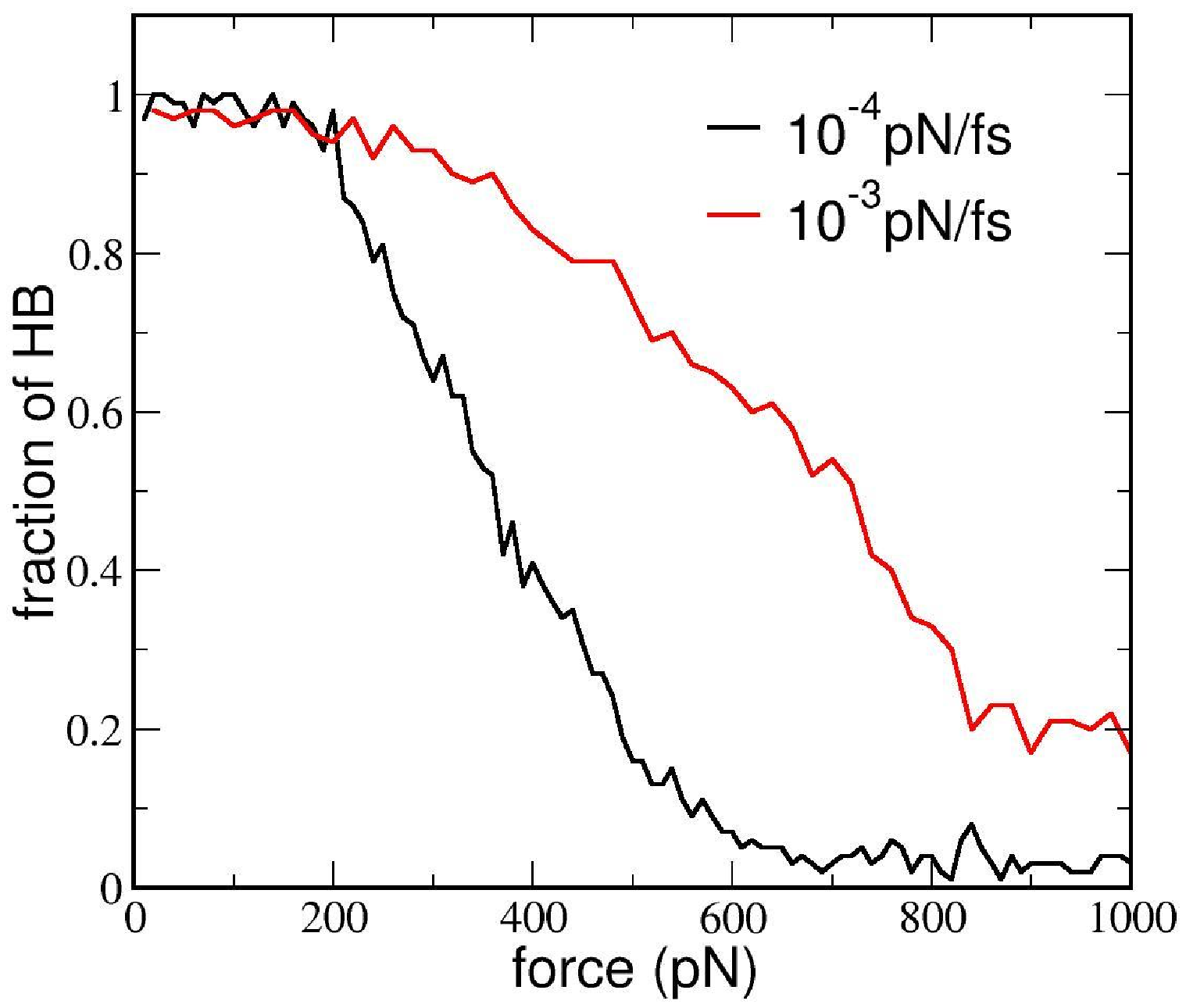}
        \label{rate-hb}
        }
	\caption{}
	\label{rate}
        \end{center}
\end{figure}

It is known from previous theoretical study that the stiffness of the polymer 
increase when pulled at faster rate \citep{evans}. So when pulling with faster 
rate the bond strength would increase dynamically and hence stretch
 modulus is expected to increase. Typical AFM pulling rates are of 
10$^4$ pN/s whereas the slowest pulling rate achieved in our simulation
 is 10$^{11}$ pN/s (i.e., 10$^{-4}$ pN/fs). Therefore due to the higher
 pulling rates employed in our simulation, the obtained stretch modulus 
is expected to be higher in magnitude compared to the results obtained
 from single molecule experiments. Similarly the magnitude of force where 
the PX/JX structure elongates roughly twice its initial contour length would
 also be more than that obtained from experiments. To see the effect of rate of 
force on force-extension behavior and stretch modulus, we have done 
simulations at two different pulling rates viz., 10$^{-4}$ pN/fs and 10$^{-3}$ pN/fs. 
Fig. \ref{rate-ext} shows the force-extension curve for PX structure with two
 different pulling rates. It is clear from the plot that the force at which 
the PX structure extends double its initial contour length is very high when 
pulled with 10$^{-3}$ pN/fs pulling rate compared to 10$^{-4}$ pN/fs pulling rate. 
This is due to the dynamic stiffening of the H-bonds in PX structure when pulled
 with 10$^{-3}$ pN/fs compared to 10$^{-4}$ pN/fs. 
{Breaking of H-bonding is the major signature of 
mechanical deformation of the DNA molecule \citep{markus_prl_2008,markus_nanolett2008}}
We have calculated the fraction of surviving H-bonds 
as a function of pulling force at two different
 pulling rates (Fig. \ref{rate-hb}). As the applied force is increased, 
the fraction of survived H-bonds decreases and at large enough force the fraction
 of survived H-bonds goes to zero. From Fig. \ref{rate-hb} it is clear that
 the fraction of surviving H-bonds goes to zero at smaller forces when
 pulled with 10$^{-4}$ pN/fs pulling rate compared to 10$^{-3}$ pN/fs pulling rate. 
The stretch modulus obtained from energy-strain and force-extension curve 
are listed in table \ref{table1} in the presence of Na$^+$ (Mg$^{2+}$) 
counterion medium.

\section{Conclusion}
\label{conclusions}
We have calculated the rigidity of PX DNA molecules in the presence of 
Na$^+$ and Mg$^{2+}$ counterions by directly calculating their 
force-extension behavior under axial stretching. Earlier we demonstrated
 \citep{maiti1,maiti2,maiti_jnn} the rigidity of these crossover DNA motifs from the
 vibrational density of states analysis but a quantitative estimate of 
their structural rigidity was lacking. Now we give a quantitative 
estimate of the stretch modulus of these DNA strictures.
 In the presence of Na$^+$ ions at a counterion concentration 
246 mM the stretch modulus of the PX-6:5 structure is almost 30 \% 
more than that of normal B-DNA double helix of same length and having
 sequence of one of the double helical domain of PX-6:5. 
The computational cost of these calculations is enormous, thus restricting 
us to use intermediate pulling rates of 10$^{-4}$ and 10$^{-3}$ pN/fs.
 To understand the effect of crossovers on the stretch modulus of these
 DNA motifs we have also calculated the force-extension profile of the
 JX$_1$/JX$_2$ motifs in the presence of Na$^+$ counterions.
 We find that JX$_1$ has stretch modulus of 1515 pN which is slightly
 smaller than the stretch modulus of PX-6:5 (1635 pN). JX$_2$ has a stretch
 modulus of 1349 pN which is 286 pN smaller than the stretch modulus of
 PX-6:5. When the DNA is pulled, among all contributions to the total 
energy of the DNA, there is a dramatic increase in torsion energy. 
This increase in torsion energy is due to a very large change in different
 torsion angles (Fig. \ref{dist_angle_torsion}) which cause the DNA to destabilize
 with increased force. Interestingly there is almost no change in 
various bond distances or angles as a function of force (Fig. \ref{dist_angle_torsion}). 
The similar behavior was observed in the presence of divalent Mg$^{2+}$ ions. 
In the presence of Mg$^{2+}$ we find the stretch modulus of PX-6:5, 
JX$_1$ and JX$_2$ to be 1875 pN, 1465 pN and 1654 pN, respectively.
 PX-6:5 has the highest stretch modulus than any other DNA motif as 
in the case of Na$^+$ medium. Interestingly all structures in 
Mg$^{2+}$ medium have more stretch modulus compared to Na$^+$ medium due
 to large electrostatic screening arising from the divalency of Mg$^{2+}$ ions.
 This could be due to the fact that presence of Mg$^{2+}$ ions gives extra 
stability to the structure, owing to the strong coordination of Mg$^{2+}$ 
with the phosphate atoms of the two double helical domains. 
We have also studied the effect of the rate of force on the rigidity of
 DNA motif and found that the increased rate of force enhances the rigidity of the structure.

\section{Acknowledgements}
We acknowledge Department of Science and Technology (DST), 
Government of India for financial support. PKM also thanks Alexander
 von Humboldt foundation for sponsoring his visit to Technical University Munich
 where part of the work was done and Roland Netz for help with the
 sequential energy fitting and valuable comments. We are also grateful 
to Prof. Ned Seeman for a critical reading of the manuscript and valuable 
suggestions. MS thanks University Grants Commission (UGC), 
India for senior research fellowship.\\

\clearpage

\bibliography{px_jx_references}
\clearpage

\begin{table}
\caption{Stretch modulus $\gamma_1$ (pN) from sequential fit to the
 energy-strain curve and polynomial fit to the force-extension curve 
in the presence of Na$^+$ (Mg$^{2+}$) counterions with different rates of force 
of 10$^{-4}$ and 10$^{-3}$ pN/fs (Fig. S15). 
Note that the values in brackets corresponds to Mg$^{2+}$ case.}
\begin{center}
	\begin{tabular}{lrrrrr}
	\hline
	& & \multicolumn{4}{c}{$\gamma_1$ for Na$^+$ (Mg$^{2+}$) (pN)}\\ 
	& & \multicolumn{2}{c}{Sequential Fit}& \multicolumn{2}{c}{Polynomial Fit}\\ 
	\cline{3-6} DNA & $c~(mM)$ & 10$^{-4}$\hspace{11 mm} & 10$^{-3}$\hspace{5 mm} & 10$^{-4}$\hspace{10.6 mm} & 10$^{-3}~~~~$\\
	 &  & (pN/fs)\hspace{6 mm} & (pN/fs) & (pN/fs) \hspace{5 mm} & (pN/fs)\\
	\hline
	B-DNA      & 254(140) & 1269\hspace{11 mm} & 1358\hspace{5 mm} & 1578(1591) & 1644\hspace{5 mm} \\ 
	PX-6:5         & 246(160) & 1636(1875) & 1737\hspace{5 mm} & 1772(1841) & 2322\hspace{5 mm} \\
	JX$_1$         & 246(160) & 1515(1465) & 1558\hspace{5 mm} & 1374(1404) & 2176\hspace{5 mm} \\
	JX$_2$         & 246(160) & 1349(1654) & 1480\hspace{5 mm} & 1521\hspace{10.6 mm} & 2051\hspace{5 mm} \\
	AT65           & 246\hspace{9 mm} & 1592\hspace{11 mm} & 1731\hspace{5 mm} & 1378\hspace{10.6 mm} & 1981\hspace{5 mm} \\
	GC65           & 246\hspace{9 mm} & 1780\hspace{11 mm} & 1862\hspace{5 mm} & 1546\hspace{10.6 mm} & 2138\hspace{5 mm} \\
        \hline	
	\end{tabular}
\end{center}
\label{table1}
\end{table}
\clearpage

\clearpage
\section*{Figure Legends}
\subsubsection*{Fig. ~\ref{init-str}.}
Atomic level structure of PX/JX DNA molecules:
(a) The basepair sequences used in the generations of
PX-6:5, JX$_1$ and JX$_2$ crossover molecules. 
Initial structure of (b) PX-6:5, (c) JX$_1$
and (d) JX$_2$ used in our pulling simulation.

\subsubsection*{Fig. ~\ref{force-ext}.}
Force-extension curves: (a) Force-extension behavior of
PX-6:5 and B-DNA molecule in the presence
of Na$^+$ counterions. (b) Force-extension behavior of 
PX-6:5, JX$_1$ and JX$_2$ in presence of Na$^+$ ions.
The rate of force is 10$^{-4}$ pN/fs.

\subsubsection*{Fig. ~\ref{dist_angle_torsion}.}
Variation of backbone parameters with force:
O$3^{\prime}$-P bond distance, P-O$5^{\prime}$-C$5^{\prime}$ angle
and C$5^{\prime}$-C$4^{\prime}$-C$3^{\prime}$ angle as a function of force
applied on PX-6:5 in Na$^+$ (black) and Mg$^{2+}$ (blue dashed line) medium.
There is very little change in the bond distance and the angles with the increase in force.
Variation of torsion angles alpha ($\alpha$), beta ($\beta$), gamma ($\gamma$),
 delta ($\delta$), epsilon ($\epsilon$) and zeta ($\zeta$) as a function of force applied
 at the end of PX-6:5 in Na$^+$ (black) and Mg$^{2+}$ (blue dashed line) medium. There is a very
 large change over 100 \% of the equilibrium zero force limits (dotted lines) in all
 the torsion angles with the force applied.

\subsubsection*{Fig. ~\ref{rate}.}
Effect of rate of pulling on PX-6:5 structure:
 (a) Force-extension curve at two different pulling rates.
 Higher the pulling rate steeper is the response and hence large stretch modulus.
(b) Fraction of survived H-bonds as a function of pulling force at
different pulling rate. At high pulling rate melting/breaking of
H-bonds occurs at larger magnitude of pulling force.

\subsubsection*{Fig. ~\ref{energy-force}.}
Energy of various PX/JX structures as a function of pulling force in 
Na$^+$ and Mg$^{2+}$ counterions. The rate of pulling is 10$^{-4}$ pN/fs.
 The minimum in the energy curve corresponds to the most stable structure during pulling.

\subsubsection*{Fig. ~\ref{snapshots-px65}.}
Instantaneous snapshots of the PX-6:5 structure at various forces
 (a) 200 pN, (b) 400 pN, (c) 600 pN, (d) 800 pN and (e) 1000 pN in the
presence of Na$^+$ counterions. {
 As the applied force is increased 
beyond 200 pN, H-bonds in the basepair were broken where the} 
extension in the DNA is almost double its initial contour length.

\begin{figure}
        \begin{center}
        \includegraphics[height=80mm]{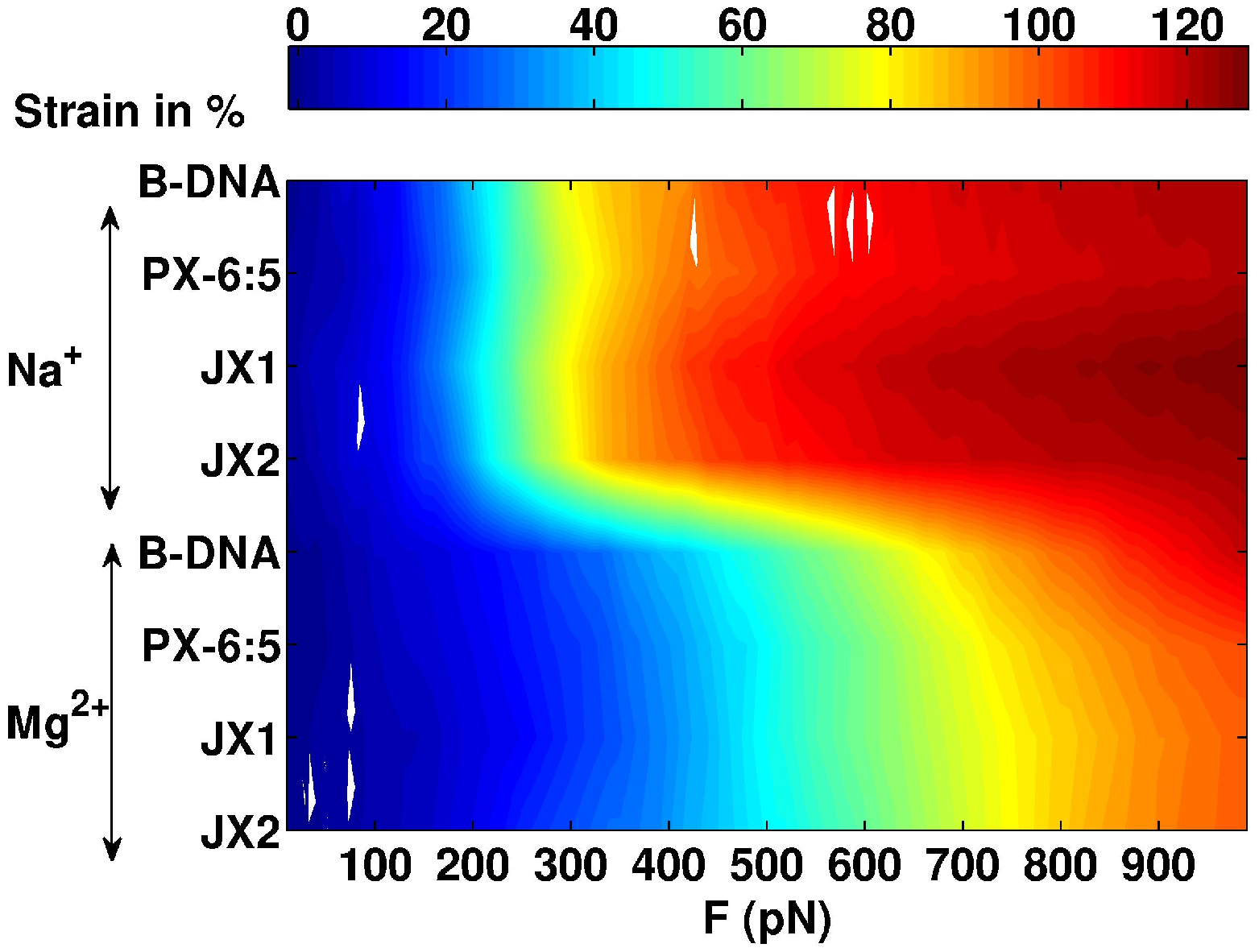}
        \caption{}
        \label{force_ext_spectrum}
        \end{center}
\end{figure}

\subsubsection*{Fig. ~\ref{force_ext_spectrum}.}
{Force-extension spectrum analysis. Upper color bar indicates
the strain increase percentage on applying the external
force on the various DNA structures. The extension in 
Mg$^{2+}$ medium is less than the extension in Na$^+$ medium 
which implies the enhanced rigidity of the PX/JX DNA molecules 
in Mg$^{2+}$ medium.}



\clearpage

\clearpage

\clearpage

\clearpage

\clearpage

\clearpage

\clearpage

\clearpage
\end{document}